# COVID-ResNet: A Deep Learning Framework for Screening of COVID19 from Radiographs

Muhammad Farooq[1] and Abdul Hafeez[2]

[1] Department of Electrical Engineering, University of Alabama, Tuscaloosa, AL, USA
Email: mfarooq@crimson.ua.edu

[2]Dept. of Computer Science & IT, University of Engineering & Technology, Peshawar, Pakistan
Email: abdul.hafeez@uetpeshawar.edu.pk

*Abstract*—In the last few months, the novel COVID-19 pandemic has spread all over the world. Due to its easy transmission, developing techniques to accurately and easily identify the presence of COVID-19 and distinguish it from other forms of flus and pneumonia is crucial. Recent research has shown that the chest X-rays of patients suffering from COVID-19 depicts certain abnormalities in the radiography. However, those approaches are closed source and not made available to the research community for re-producibility and gaining a deeper insight. The goal of this work is to build open source and open access datasets and present an accurate Convolutional Neural Network framework for differentiating COVID-19 cases from other pneumonia cases. Our work utilizes state of the art training techniques including progressive resizing, cyclical learning rate finding and discriminative learning rates to training fast and accurate residual neural networks. Using these techniques, we showed state of the art results on the open access COVID-19 dataset. This work presents a 3-step technique to fine-tune a pre-trained ResNet-50 architecture to improve model performance and reduce training time – we call it COVID-ResNet. This is achieved through progressively re-sizing of input images to 128x128x3, 224x224x3, and 229x229x3 pixels and fine-tuning the network at each stage. This approach along with the automatic learning rate selection enabled us to achieve state of the accuracy of 96.23% (on all the classes) on the COVIDx dataset with only 41 epochs. This work presented a computationally efficient and highly accurate model for multi-class classification of three different infection types from along with Normal individuals. This model can help in early screening of COVID-19 cases and help reduce burden on healthcare systems.*

*Keywords—COVID-91, Corona Virus, Deep Learning, COVIDx, Classification, ResNet*

## I. INTRODUCTION

The COVID-19 is the result of the infection of individual caused by the acute respiratory syndrome coronavirus (SARS-CoV-2). This pandemic is spreading all over the world with recorded rate that is never seen before for any infectious disease. This can be spread even by individuals who are asymptomatic. One of the effective techniques proposed by World Health Organization (WHO) to control the spread of the viral infection is social distancing and contact tracing. Therefore, a critical step in this direction is an effective and accurate screening of the COVID-19 patients for not only receiving quick treatment but also isolation from the public to halt spreading of the viral infection. State-of-the-art techniques for the detection of COVID-19 and those that measure the production of antibodies in response to the infection caused by the said virus include serology and reverse transcription polymerase chain reaction i.e., rRT-PCR [1]. Clinical setups and population surveillance employ serology for the detection of antibodies. The limited availability of the test kits makes it challenging to detect every individual affected by the virus. Furthermore, these tests take from few hours to a day or two to produce the output, which becomes too tedious, time consuming and most of the time error prone in the current state of emergency. Therefore, a faster and reliable screening techniques that could be further confirmed by the PCR test is urgently required.

Some studies have shown the use of imaging techniques such as X-rays or Computed Tomography (CT-scans) for finding characteristic symptoms of the novel corona virus in these imaging techniques [2], [3]. Recent studies suggest the use of chest radiography in the epidemic areas for the initial screening of COVID-19 [4]**.** Therefore, the screening of radiography images can be used as an alternate to the PCR method, which exhibit higher sensitivity in some cases [5]. Nevertheless, the main bottleneck that the radiologists experience in analyzing radiography images is the visual scanning of the subtle insights. This entails the use of intelligent approaches that can automatically extract useful insights from the chest X-rays those are characteristics of COVID-19.

A number of studies have shown the ability of neural networks, especially convolutional neural networks to accurately detect the presence of COVID-19 from CT-scans [2], [6]. However, the datasets are often not publicly available, which reduces their access to the wider research community and further development of classification techniques on standardized datasets. This works builds on the research presented in [7], in hopes of building open source and open access techniques to help in the fight against the COVID-19 pandemic. Authors in [7] presented a dataset that they called COVIDx and a neural network architecture called COVID-Net. The dataset consists of chest radiography images belonging to 4 classes i.e. Normal comprising cases without any infections, Bacterial pneumonia, Viral pertaining to non-COVID-19 pneumonia, and COVID-19. They reported an overall accuracy of 83.5% for these four classes. Their lowest reported positive predictive value was for non-COVID-19 class (67.0%) and highest was for Normal class (95.1%). We are using the same dataset with the following goals:

1) Improve the overall accuracy of the model for all the four classes with positive predictive values > 90 (among all classes).
2) Use network architecture with lower number of parameters and computational needs. An architecture that provide a balance between performance and computational complexity.
3) Use techniques for training models which need a lower number of epochs and hence faster training.

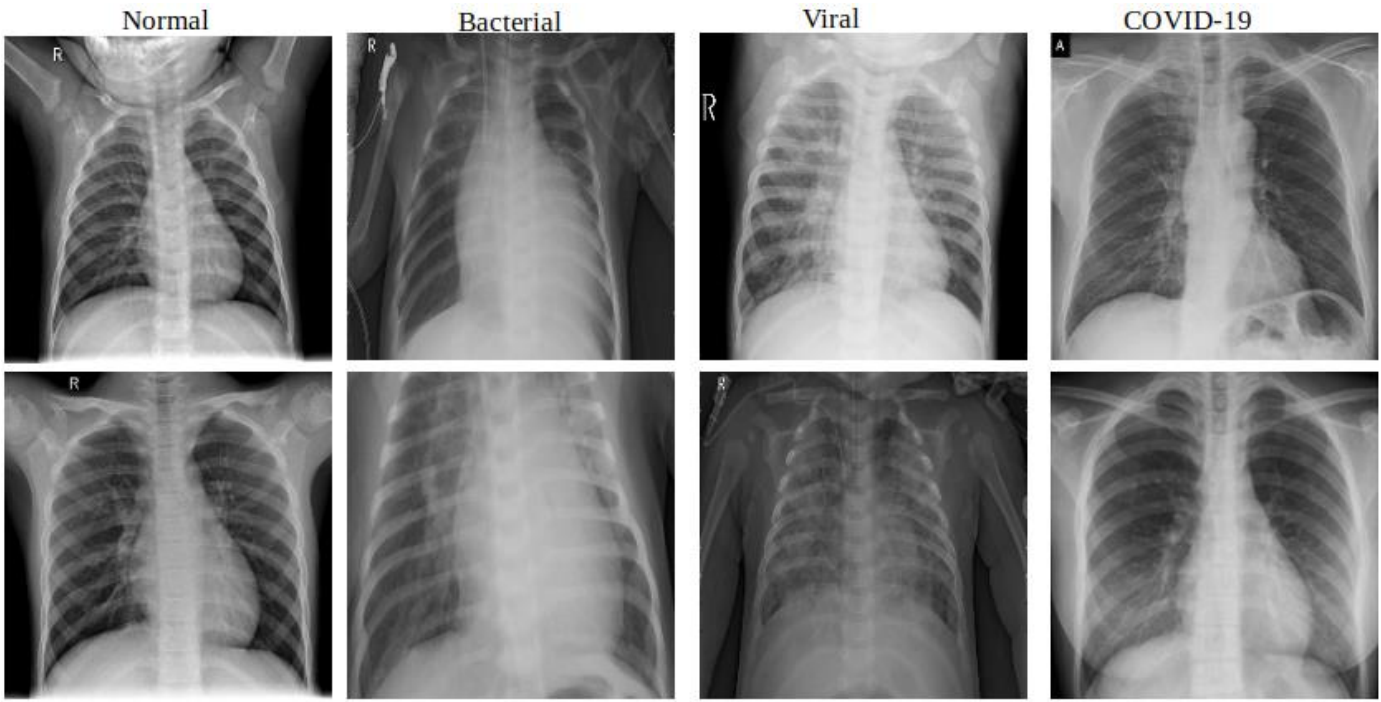

Figure 1. Example chest radiographs of four classes present in the COVIDx dataset

We utilize state of the art techniques to achieve these goals with continuous human input and show that human intervention in the training can significantly improve both the performance of the models and reduce training time. This work utilizes progressive image re-sizing along with automatic learning rate finding [8] and discriminative learning rate [9] to achieve state of the art results on this task.

The rest of the paper is organized as follows: Section II captures the Methods that provide details of the dataset and the proposed architecture. Section III demonstrates the Results. Discussion is illustrated in Section IV and Section V concludes the paper.

II. METHODS

*A. COVIDx Dataset*

We used the COVIDx dataset that was recently made public by the authors of the COVID-Net [7]. It consists of a total of 5941 posteroanterior chest radiography images from 2839 patients with 4 classes namely 1) Normal (no infections), 2) Bacterial (bacterial pneumonia) 3) Viral (non-COVID-19 viral pneumonia) 4) COVID-19. The dataset was curated by combining two publicly available datasets. The authors have made a pre-processed version of the dataset available at https://github.com/lindawangg/COVID-Net.

In the current version of the dataset, there are 68 COVID-19 radiographs from 45 COVID-19 patients. There were a total of 1203 patients with negative pneumonia (i.e. Normal class), 931 patients with a bacterial pneumonia and 660 patients with non-COVID-19 viral pneumonia cases. As observed, the COVID-19 cases are significantly lower than the other classes and we are faced with an imbalance classification problem. Details of the train and test distribution as well as the distribution of patients are given in [7] and those were used in the current study. Figure 1 shows random examples of the radiographs for all 4 classes. Distribution of the training data examples and the number of patients' distribution is shown in Figure 2 and 3 respectively. Please refer to [7] for further details.

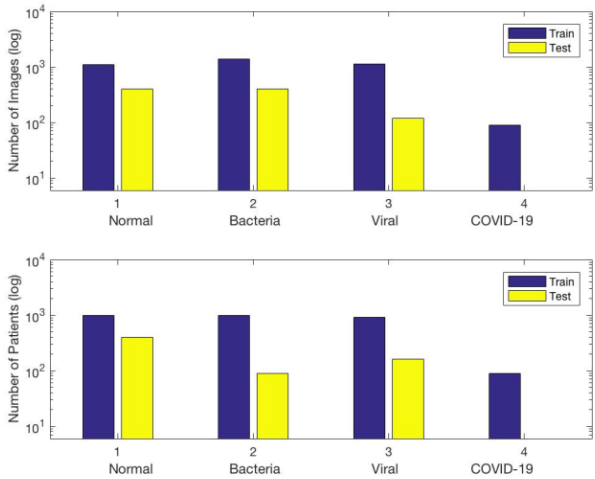

Figure 2 The upper graph shows the distribution of images for individual infection type of the COVIDx dataset. Normal refers to negative infection. The blue bar represents the number of training data, while yellow bar represents the aggregate of the test set. The below sub graph represents the aggregate of the patients in each category of the COVIDx dataset.

*B. Data Augmentation*

Data augmentation helps in creating newer examples by applying different transformation randomly to the available training images. In this work, the transformation that we used included vertical flips of the training images, random rotation of the images (maximum rotation angle was 15 degree), and lighting conditions. Data augmentation increases the size of input training data along with the model regularization and hence improving generalization of the training model. Only the training data was augmented, and the test time augmentation was not explored.

*C. Network Architecture*

Rather than proposing our own architecture, we have leveraged the knowledge from a pool of the already existing Convolution Neural Network architectures that have shown excellent results using a wide variety of classification tasks. We employ a variant of the residual neural network with a total of 50 layers called ResNet50 [10]. Residual Networks provide a good combination of performance and number of parameters and have proved faster training. Another reason for using the residual network architecture is the ability to feed images of sizes other than with which they are trained with. This is a critical part of the training methodology employed for training a high-performance network with very few epochs using the techniques introduced in fastai [11]. We call the resultant network as COVID-ResNet.

The weights used in ResNet50 are pretrained with ImageNet dataset [12]. In COVIDx dataset, the authors have rescaled all images to 224x224x3. Those images are further rescaled to 128x128x3, 224x224x3 and 299x299x3 and employed in different stages of training (*Subsection D*). The images are normalized using the per channel mean and standard deviation of the images present in the ImageNet dataset. This is critical as we are using the pretrained weights of the network that is previously trained with ImageNet dataset.

For transfer learning, the head of the trained model is replaced by another head containing a sequence of Adaptive average/max pooling, batch normalization, drop out and linear layers as proposed in [11]. This network is further fine-tuned with the COVIDx dataset as described below.

*D. Training the Network*

Progressive resizing is used where the network is initially fine-tuned with smaller images and the size of the input images is gradually increased as the training progresses. This can be done as the features learned by the different layers of the CNN are independent of the input image size. Furthermore, the global features are still present in the same resized image with different pixel resolutions. Training is performed in three stages where each stage corresponds to images of different input dimensions. In each stage, instead of manually tweaking the learning rates, we use the Cyclical Learning Rate technique proposed by Leslie Smith in [8] for help selecting optimal learning rate.

*Stage - I:* The input images are resized to 128x128x3 pixels and the COVID-ResNet are tuned in 2 steps. 1) only the newly added head of the network is trained while preserving the ImageNet weights for the rest of the body with a learning rate of 1e-3 for 3 epochs. 2) the whole network is fine-tuned (both the body and the head of the model) using discriminative learning rate proposed in [9] for 5 epochs.
*Stage - II:* The head of the model resulting from stage-I is further fine-tuned with images of size 224x224x3 in the first step with learning rate of 1e-4 for 3 epochs. In the second step, the whole network is further fine-tuned for 5 epochs with discriminative learning rate as before.
*Stage - III:* In the last stage, the whole network is further fine-tuned with input images of size 229x229x3 for 25 epochs. In this case we use discriminative learning rates where the earliest layer was trained with a learning rate of 1e-6 and the last layer was trained with a learning rate of 1e-4. All the layers in between were trained with equidistance learning rates between these two values.

Training the model in multiple stages with different input image sizes using progressive resizing tends to achieve a better generalization. This is also an example of transfer learning from one input image size to another. One thing to notice is that as we proceed to the later stages of the training, we ensure to reduce the learning rates. This ascertains that the weights are not modified much from one stage to successive stage. Adam optimizer with batch size of 32 is used in training. The entire data preprocessing, data augmentation and training performed use the fastai [11] framework.

III. RESULTS AND DISCUSSION

In this section, we present the results of the network performance and the quantitative metrics that are used. For each infection type as well as the normal/healthy cases, we present the sensitivity (recall), positive predictive precision (positive predictive values), and the F1-score. Table 1. presents the number of total parameters of the network (both trainable and non-trainable), total number of training epochs and the accuracy of the model on independent test set as suggested by authors in [7]. Compared to the original results on the COVIDx dataset, we have achieved a significant performance improvement of about 13% (96.23% compared to 83.5% in COVID-Net) with 4.5 times lower parameters (25.6M vs. 116.6M COVID-Net).

Table. 2, 3 and 4 presents the sensitivity (recall), Positive Predictive Value (PPV) and the F1-scores for each class and the overall confusion matrix is presented in Figure 4.

Table 1. Performance of COVID-ResNet on COVIDx test dataset

| Comparison | Params (M) | Training Epochs | Accuracy (%) |
|---|---|---|---|
| COVID-ResNet | 25.6 | 41 | 96.23 |
| COVID-Net | 116.6 | 100 | 83.5 |

Table 2. Sensitivity of each class

| Recall (Sensitivity) % | | | |
|---|---|---|---|
| Normal | Bacterial | Viral | COVID-19 |
| 96.58 | 97.15 | 93.96 | 100.0 |

Table 3. Precision (Positive Predictive Value) of each class

| Positive Predictive Value (Precision) % | | | |
|---|---|---|---|
| Normal | Bacterial | Viral | COVID-19 |
| 99.12 | 95.60 | 92.72 | 100.0 |

Table 3. F1-score of each class

| F-1 score % | | | |
|---|---|---|---|
| Normal | Bacterial | Viral | COVID-19 |
| 97.84 | 96.37 | 93.33 | 100.0 |

From Table 2, we can see that COVID-ResNet achieves a perfect sensitivity for the COVID-19 class i.e., it did not miss any of the cases of the COVID-19. This is critical as the goal is to be able to detect all positive COVID-19 cases to reduce the community spread. Another important aspect of the results is that the network showed a stronger positive predictive value of 100%. This shows that there were no classes falsely misclassified as COVID-19 from the other classes (infections).

XXX-X-XXXX-XXXX-X/XX/$XX.00 ©20XX IEEE

Identify applicable funding agency here. If none, delete this text box.

Higher PPV will ensure eliminating the high-demand PCR tests for non-COVID-19 cases. A similar trend can be seen in terms of F1-score. Although the results for COVID-19 are perfect for this test dataset (all metrics are 100%), we want to emphasize that the nature of the very limited test set. The number of test cases for COVID-19 are very small compared to the other classes. We plan to test and improve our model as additional COVID-19 data becomes available.

Compared to the results shown in [7], we have shown significant performance improvements for the COVID-19 as well as other classes. Out of the 637 test cases, only 24 cases are miss classified. Similar to [7], we also observe that the non-COVID-19 viral infection class has lower PPV compared to the other classes. In contrast to [7], we employ a data augmentation relevant to the our data that has resulted in a significant performance improvements.

One of the very encouraging results is the ability of the model to achieve higher sensitivity and PPV on the Normal class. This will help in ensuring that there are no false positive cases not only for the COVID-19 but also for the other two infection classes and help alleviate the burden on the healthcare system. We further plan to add techniques to explain the model prediction as to why certain decisions were made to better understand the rationale of the decision-making process. This will also add more confidence in the model results as well as the ability to find newer interesting insights and patterns in the radiographs.

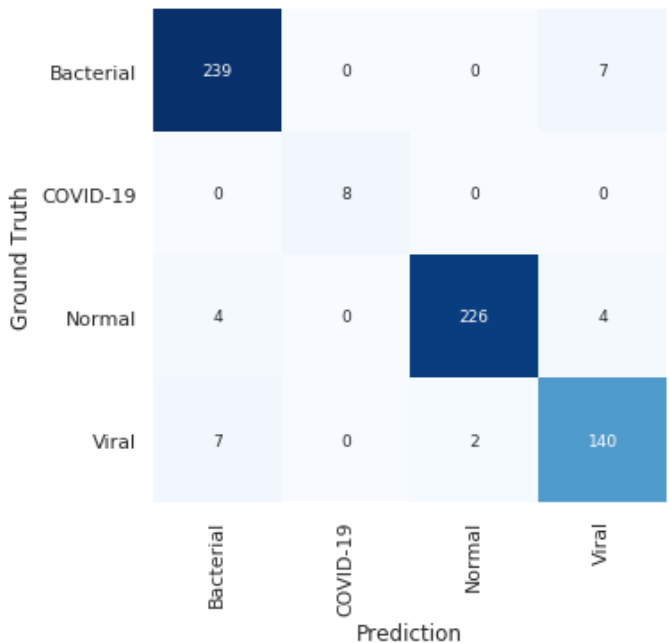

Figure 4: Confusion matrix for COVID-ResNet on the COVIDx test dataset

## IV. CONCLUSIONS

In this work, we present COVID-ResNet for classification of COVID-19 and three other infection types. COVID-ResNet was trained on a publicly available dataset COVIDx and have shown excellent classification accuracy across all the classes on the independent test dataset. We also showed the importance of data augmentation in order to increase the training set size and improve generalization. We showed that using state-of-the-art techniques along with human intervention during training can help finding appropriate learning rates for an improved performance and speed.

We would like to emphasize that even though COVID-ResNet is very promising and accurate, it is not meant to be directly employed for clinical diagnosis. The goal of this work was to show that using different training techniques enable us to train models that are computationally efficient and accurate. In order to make COVID-ResNet clinically useful requires training with a larger dataset and testing in the wild with a larger cohort.